\documentclass[conference]{IEEEtran}
\usepackage[utf8]{inputenc} 
\usepackage{longtable}
\usepackage{booktabs}
\usepackage{etoolbox}
\usepackage{needspace}
\usepackage{cite}
\usepackage{amsmath,amssymb,amsfonts}
\usepackage{mathabx}
\usepackage{algorithmic}
\usepackage{subcaption}
\usepackage{graphicx}
\usepackage{textcomp}
\usepackage{xcolor}
\usepackage{float} 
\usepackage[font={small}]{caption}

\newcommand{\curlb}[1]{\left\{#1\right\}}
\newcommand{\squareb}[1]{\left[#1\right]}

\newcommand{\Tau}{\mathrm{T}}
\begin{document}
\title{Discrete-Space Generative AI Pipeline for Semantic Transmission of Signals} 

\author{
\IEEEauthorblockN{\normalsize Silvija Kokalj-Filipovic, Yagna Kaasaragadda}
\IEEEauthorblockA{\small Rowan University\\
\small\em kokaljfilipovic@rowan.edu, yagna@tror.ai}}
\maketitle
\begin{abstract}
We introduce  {\bf Discernment}, a semantic communication system that transmits the meaning of physical signals (baseband radio and audio) over a technical channel using GenAI models operating in discrete spaces. Discernment dynamically adapts to channel impairments—modeled as erasure channels—by switching between an autoregressive or a diffusion-based generative algorithm, depending on the erasure pattern. Our results show that Discernment maintains semantic integrity even as channel capacity severely degrades, exhibiting very small and graceful performance decline in both classification accuracy and statistical fidelity of the reconstructed meaning.  These findings demonstrate Discernment’s ability to adjust to diverse physical channel conditions while maintaining spectral efficiency and low model complexity, making it well suited for IoT deployments and strongly motivating further research on this semantic-channel paradigm.
\end{abstract}
\vspace{-2mm}
\section{Introduction}
\label{introduction}
Semantic communication \cite{semantic} is a new concept, emerging in the context of generative AI ({\em GenAI}): to convey the meaning of a message, both communicating sides instantiate an identical generative model that extracts the meaning from a compressed data representation, informed by a task objective. This is in contrast with the classical communication framework, in which we aim to efficiently send any bit-encoded message over a physical medium with a low bit error rate. Classical communication models are unaware of the meaning aligned with the objective of the task utilizing the message, and therefore bits of no use to the task may be sent. As semantic communication is gaining traction, one of the limitations is that use cases are tied to particular application domains such as vision, speech, and text \cite{aishwarya2024generative}, while semantically transmitting a wide variety of signals remains unexplored, leading to a lack of insight and generalization. Further, generative models that enable semantic communication are computationally complex and power-hungry. Exploring semantic communications in the discrete latent domain may alleviate this problem. 
We study a semantic communication framework which leverages discrete space latent representations. Our approach models the technical communication channel as an erasure channel and provides generative solutions to recover from different erasure patterns. It includes the steps of (a) learning discrete latent representations of data, optimized for various tasks (b) using the resulting arrays of discrete tokens to train generative models (GenAI)  (c) communicating only a subset of tokens to the objective task, and letting the trained GenAI model at the destination  infer the rest of the tokens (erasures) to perform information recovery for the completion of the objective task. (d) matching the GenAI type to the expected patterns of erasures in the technical channel.

The proposed generative framework is called Discernment, a wordplay that emphasizes both the discrete-space  and perception. Discernment is computationally efficient yet highly effective for signal synthesis, supporting both data augmentation and semantic communication. We develop and evaluate two model families: (a) transformer-based architectures operating in discrete latent spaces and (b) a Score Entropy Discrete Diffusion (SEDD) model.  SEDD  \cite{lou2024discretediffusionmodelingestimating} is a recent discrete diffusion approach to language generation, which is here adapted for  non-autoregressive  generation of diverse signals. Compared to transformers, SEDD  improves the latency of generative synthesis and reduces the model complexity. It is also suitable for random erasure patterns. Using these architectures, we demonstrate the generation of synthetic radio and audio signals as well as efficient task-aligned semantic communication. While classification alone does not justify semantic communication—since a transmitter could simply classify the signal and send the label—we use classification as a convenient and generalizable means to illustrate Discernment’s effectiveness in task-aligned semantic communication.
Our models are lightweight, making them suitable for IoT devices that must semantically convey locally acquired signals to more capable platforms for downstream processing.
The paper is organized as follows: Section~\ref{sec:technicalbackground} introduces foundational elements of the Discernment framework. Section~\ref{sec:methodology} describes the detailed methodology for signal modeling and generation. Section \ref{sec:experiments} illustrates how Discerment works in a semantic communication system and how it can respond to the  specifics of the technical communication channel. Section \ref{sec:conclude} concludes with insights for future research.
\vspace{-2mm}
\section{Technical Background}
\label{sec:technicalbackground}
This section introduces the foundational technologies that form the basis of Discernment, including discrete representation learning, transformer architectures, diffusion models, and their application for sematic communication of signals from the audio and radio-frequency (RF) domains.
\vspace{-1mm}
\subsection{Vector Quantized Variational Autoencoders (VQVAE)}
Vector Quantized Variational Autoencoders (VQVAE), introduced by van den Oord et al. \cite{DBLP:journals/corr/abs-1711-00937}, provide a mechanism for learning discrete latent representations by combining vector quantization and autoencoders via variational inference. Unlike traditional VAEs, which operate in continuous latent spaces, VQVAE learns a codebook of  embeddings. The encoder stochastically maps an input signal to the nearest codeword in the codebook, and the decoder reconstructs the input from this quantized representation.
This discrete representation simplifies the generative modeling task and reduces memory and compute requirements, especially important for modeling high-dimensional signals like RF and audio waveforms and spectrograms.
In Discerment, VQVAE is the first stage of the generative pipeline, mapping real valued signals into sequences of codebook indices, whose joint probability is later modeled using autoregressive or diffusion-based techniques.
\vspace{-2mm}
\subsection{Transformer Architectures}
Transformers, introduced by Vaswani et al. \cite{vaswani2023attentionneed}, utilize self-attention to model dependencies across the entire input sequence.  Decoder-only transformers (DoT), such as GPT-style models, are well-suited for autoregressive generative tasks, predicting the next token in a sequence conditioned on previous tokens.
We use lightweight versions of DoT models (NanoGPT \cite{Karpathy2022} and MONAI \cite{monai2024}) to model the distribution over the latent sequence of codebook indices $z_Q$ as our signal inputs are not long. These models process sequences of tokens (produced by the VQVAE encoder) and learn to generate most probable  continuations. 
\vspace{-2mm}
\subsection{Discrete Diffusion Models}
Denoising diffusion models generate samples by learning to reverse a forward process that incrementally adds noise. While most early diffusion models worked on continuous data, discrete variants such as D3PM \cite{austin2023structureddenoisingdiffusionmodels} extend this idea to categorical sequences.

The Score Entropy Discrete Diffusion (SEDD) model \cite{lou2024discretediffusionmodelingestimating} applies this principle to language generation using transformer encoders. Before training, input sequences are corrupted by masking a subset of tokens; the model learns to reconstruct the original sequence. Unlike autoregressive models, diffusion models estimates all tokens in parallel. 
In this work, SEDD is applied for the first time to signals other than language. SEDD complements DoT models by offering a non-autoregressive alternative for sequence generation.
\vspace{-2mm}
\subsection{Generative Modeling in RF and Audio Domains}
Generative modeling has recently expanded into signal processing. In speech, VQVAE based systems have been used for high-fidelity synthesis, as shown in Jukebox \cite{dhariwal2020jukebox} and other approaches. However, models like HuBERT \cite{hsu2021hubert} and Jukebox \cite{dhariwal2020jukebox} are highly complex, combining multi-level quantization, adversarial loss terms, and hybrid transformer-GAN architectures (VALL-E and EnCodec \cite{defossez2022high}). Efficient  learning of discrete-space priors is also demonstrated with simple architectures like VQalAttent \cite{rodriguez2024vqalattent}, which is convenient in the context of semantic channels.  Among many examples of DoT models trained for signal generation, we mention several in the audio and radio domains: AudioLM \cite{borsos2023audiolm} and Conformer \cite{xu2022conformer}, demonstrated the potential of transformers in audio modeling; Likewise, ReFormer \cite{ReFormer} demonstrated that transfomers  trained on VQVAE indices can generate high fidelity digitally-modulated RF signals. 
RF-Diffusion \cite{chi2024rfdiffusionradiosignalgeneration} uses continuous-space diffusion on radio signals modeled in the time-frequency domain.
We present the first comparative analysis of transformer- and diffusion-based generative models operating on discrete latent representations across multiple signal types. The analysis evaluates classification-aligned synthetic signal quality, computational complexity (Table~\ref{tab:configs}), and the effectiveness in semantic communication systems. This multi-domain evaluation of fidelity, diversity, and task alignment, particularly when synthesis is conditioned on channel erasure patterns, is necessary for generalizing semantic communication scenarios. 
\vspace{-2mm}
\begin{table}[!htbp]
\centering
\caption{{Configurations Used to Evaluate Generative Models}}
\begin{tabular}{lccccc}
\toprule
\textbf{-} & \multicolumn{1}{c}{\textbf{Codebook}} & \multicolumn{1}{c}{\textbf{VQVAE}} & \multicolumn{1}{c}{\textbf{Monai}} & \multicolumn{1}{c}{\textbf{NanoGpt}} & \multicolumn{1}{c}{\textbf{SEDD}} \\
 & \textbf{Size} & \textbf{Params} & \textbf{Params} & \textbf{Params} & \textbf{Params} \\
\midrule
Conf. 1 & 128  & 331K  & 1.30M  & 1.28M  & 2.02M \\
Conf. 2 & 256  & 347K  & 5.06M  & 5.00M  & 2.06M \\
Conf. 3 & 512  & 380K  & 19.97M & 19.70M & 2.12M \\
Conf. 4 & 1024 & 446K  & 79.26M & 78.19M & 2.25M \\
\bottomrule
\label{tab:configs}
\end{tabular}
\end{table}
\vspace{-7mm}
\subsection{Semantic Communication System Model}
A convenient information-theoretic definition of a semantic communication (SC) channel was recently formulated in \cite{semanticTheory} given the generative knowledge of the “language” $W$ and its interpretation of the meaning $w\in W$ of a message $s.$ The authors define the language as a mapping that relates a message $s$ to the meaning $w: q(w\mid s) \in \squareb{0,1}, w\in W,s \in S,\ \sum_w{q(w\mid s)}=1$. Consequently, the semantic channel is defined as the probability of the communicated meaning $\hat{w}$ given the intended meaning $w$:
\begin{equation}Pr( \hat{w}\mid w)=\sum_{s,\hat{s}}{p(s \mid w)c(\hat{s}\mid s)q(\hat{w}\mid \hat{s})}, \vspace{-2mm}    
\end{equation}
where $\hat{s}$  is the received version of the message $s$ transmitted over the technical channel  $c(\hat{s}\mid s)$.  Obviously, the quality of the semantic channel is determined by the manner in which technical communication is implemented (e.g., coding and modulation schemes, etc.). As Discernment addresses semantic communication in the domain of discrete tokens, we utilize the erasure channel model to represent technical communication. The erasure channel effectively models both truncation and puncturing of the latent array $z_Q$. {\bf Truncation} can occur when the receiver imposes strict latency constraints that prevent full sequence delivery over a slow channel. {\bf Puncturing} originates from packet loss, where $z_Q$ is received with missing elements. In our framework, we use different designs of trained GenAI to recover from erasures, depending on their pattern. To recover a truncated  $z_Q,$  we use a DoT as it is trained to generate tokens autoregressively, and to recover a punctured  $z_Q,$ we utilize SEDD whose inference (and training) is based on denoising a masked $z_Q$. Without GenAI modules operating at both sides of the semantic channel, the system would be unable to recover from erasures in a way that preserves task performance.  VQVAE decoder alone can perform the reconstruction task objective only if all $d_s$ tokens of the $z_Q$ are received. Thus, Discernment operates as a jointly source–channel encoded semantic communication system. For clarity and brevity, we do not present additional formal definitions and theorems, even though a more detailed treatment—-including precise erasure-channel models tailored for our use cases—-would be desirable.
\vspace{-4mm}
\section{Methodology}
\label{sec:methodology}
In this section, we describe the detailed methodology for signal dataset modeling and generation. The work involves utilizing two major datasets: AudioMNIST \cite{BECKER2024418} and TorchSig \cite{boegner2022largescaleradiofrequency}.  For either dataset, we  train a series of deep learning architectures, including VQVAE, DoT, and SEDD, for both the conditional and unconditional signal generation. Each step is carefully designed to maximize efficiency while maintaining the realism of the generated samples.
The essence of generative models is to approximate
the distribution of real data, enabling the generation and completion of data
consistent with the training set. 
\vspace{-2mm}
\subsection{Datasets}
\subsubsection{RF Modulated Signal} 
Note that this dataset models the use cases in which the radio is not the information carrier but the information itself. Instead of IQ samples being immediately demodulated to recover user data,  they are treated as  raw data for machine learning inference, signal analysis, and network intelligence applications at the edge of the Open RAN architecture. 
The process of embedding digital or analog information onto a high-frequency carrier wave is known as modulation. In digital communications, modulation involves encoding discrete bit sequences into physical signal parameters. Common digital modulation schemes from \cite{boegner2022largescaleradiofrequency} include Amplitude Shift Keying (ASK), Frequency Shift Keying (FSK), Phase Shift Keying (PSK), and Quadrature Amplitude Modulation (QAM), but more complex radio waveforms exist that combine different schemes (e.g., OFDM). To enable higher spectral efficiency and facilitate complex modulation schemes, modern digital communication systems often adopt In-phase (I) and Quadrature (Q) components in signal representation. This approach decomposes the signal into two orthogonal parts: $s(t) = I(t) cos(2\pi ft) - Q(t) sin(2\pi ft ),$
where $f$ is the carrier frequency, $I(t)$ is the in-phase component aligned with the cosine reference, and $Q(t)$ is the quadrature component, shifted 90 degrees out of phase. This formulation permits the simultaneous modulation of both amplitude and phase.
These components are typically represented as complex samples known as IQ samples: $h(t) = I(t) + j Q(t)$
where $j = \sqrt{-1}$ denotes the imaginary unit, and $h(t)$ is known as baseband signal.  To prepare a dataset, a modulated signal
$u$ is obtained as $u = M_a(b)$, where $a \in \mathcal{A}$ is the employed
digital modulation scheme. For any $a$, $M_a = \curlb{0,1}^{m(a)} \rightarrow \mathcal{C}^v$ describes the modulation function which encodes random bit sequence $b$ of length $m$ into a sequence $u$ of complex valued numbers of length $v.$ We create datapoints as sub-sequences  $x$ of $u \in \mathcal{C}^v,$ of length $p = 1024,$ and implement  the training dataset $X_{train}$ of 6 modulation classes 
\vspace{-1.5mm}
\[
\mathcal{A} = \{ \text{4ASK}, \text{8PAM}, \text{16PSK}, \text{32QAM-X}, \text{2FSK}, \text{OFDM256} \}.
\] by using an open-source library {\em torchsig} featured in \cite{boegner2022largescaleradiofrequency}. $X_{train}$ contains RF samples of high SNR, for the sake  of simplicity and  to match our Open RAN use case. The torchsig function {\em ComplexTo2D} is used to transform vectors of complex-valued numbers into  2-channel datapoints,  referred to as the {\em I} and {\em Q} channel.
\subsubsection{Audio Data} 
Our audio data is based on the AudioMNIST dataset, 
which comprises short audio 
samples of spoken numerals (0-9) \cite{BECKER2024418}. 
AudioMNIST datapoints are recordings of a variety of speakers 
of different ethnicities and genders.
They are preprocessed by 
resampling at 22,050 Hz, trimming the leading and trailing 
silence and adjusting the duration to one second by occasionally appending zeros. We apply the short-time Fourier transform (STFT) to each preprocessed datapoint, which produces a complex-valued spectrogram, represented as a tensor $\mathbf{s}$ with $F=128$ complex-valued Fourier coefficients per temporal frame $\tau\in\squareb{1,\cdots, \Tau=44}$. Finally, a complex-to-2-channel transform is applied to each $\mathbf{s}$  producing datapoints that are 2-channel real-valued  spectrograms. 
\subsection{VQVAE: Discretization of Input Space}
\begin{figure}[!htbp]
    \caption{VQVAE Architecture with Codebook $Q$ of $N=64$ codewords}
    \centering
    \includegraphics[width = 3.5in]{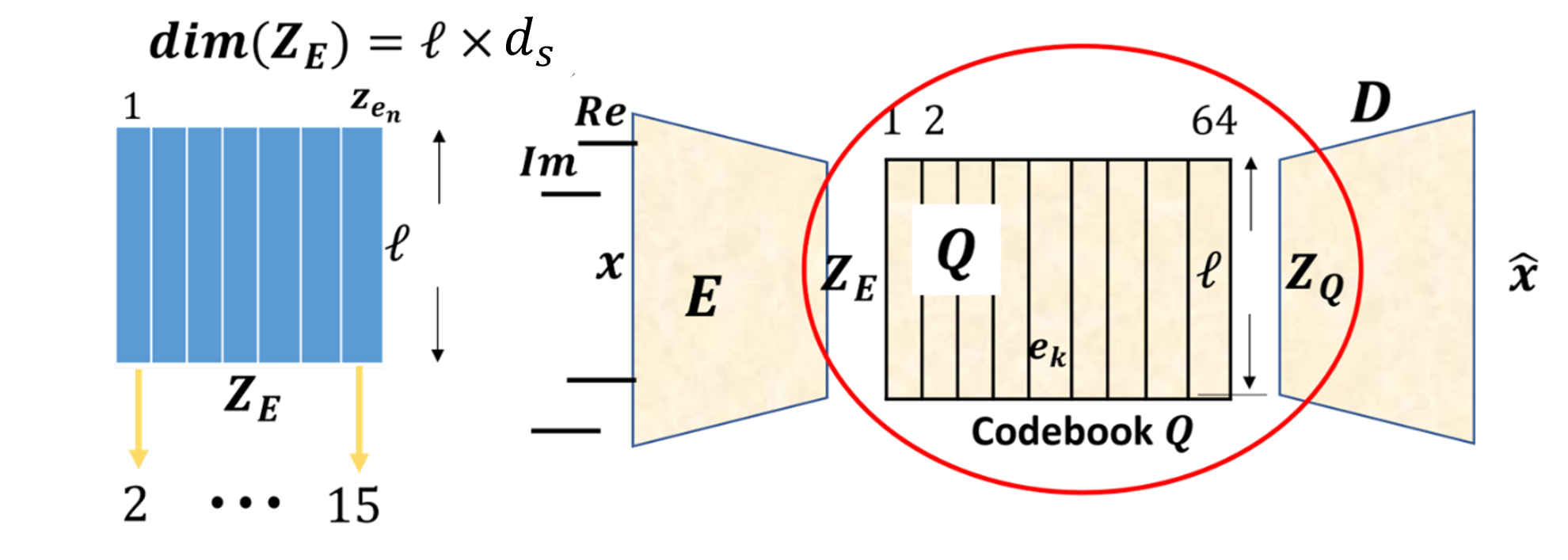}
    \label{fig:VQVAE Architecture} \vspace{-5mm}
\end{figure}

 We use a VQVAE, shown in Fig.~\ref{fig:VQVAE Architecture}, to discretize  high-dimensional Torchsig input space $X \in {\mathbb{R}}^{2\times p},$ or AudioMNIST spectrograms $X \in {\mathbb{R}}^{2\times\Tau\times F},$ into  latent tokens ${z_Q} \in \{0,1,2,…,{N-1}\}^{d_s}$ , enabling efficient sequence modeling. Here, each token in $z_Q$ is an integer index corresponding to a codeword, taking values in the range $N_Q=\{0,1,2,…,{N-1}\}.$ 
 The encoder $E_\theta$ comprises blocks of convolutional layers with residual connections. It reduces the input resolution before projecting it into a trainable codebook $Q$. Each output slice $z_e$ of $z_E$ is quantized to its nearest codeword $e_k$ from $Q$. Quantization can either be deterministic (nearest neighbor) or stochastic. Stochastic quantization is a regularization mechanism to minimize the probability of mode collapse.
 
To obtain $z_Q$, each output slice $z_e$ is compared to all codewords $e_j \in Q$, and the codeword with index $j=k$ is selected based on distance. This index $k$ is a discrete token $z_Q^i$. The full latent tensor $z_Q$ is a sequence of $d_s$ such tokens. The length $d_s$ depends on the codebook design $Q\in \mathbb{R}^{\ell\times N}$, since  $d_s=K/\ell,\ K=prod(dim(Z_E))=512\times128.$   For the Torchsig design in Conf.2 (Table~\ref{tab:configs}), 
$N=256$ codewords of length $\ell=128$ are used, i.e, $z_Q$ is a sequence of 
$d_s=K/\ell=512$ discrete indices  $k,\ 1\leq k \leq 256$. 

Note that Torchsig and AudioMNIST designs differ in both the  input dimension, and the parameter $K.$  
The decoder $D_\theta$ mirrors the encoder's structure using up-sampling and transposed convolutions to reconstruct the input signal from the quantized embeddings, which are looked up from the codebook using the stored token indices. The choice of codebook size $N$, codeword dimension $\ell$, and depth of the encoder-decoder networks are critical for balancing reconstruction quality and compression rate. 
    
The total loss of VQVAE addresses reconstruction fidelity $L_{\text{recon}}$, codebook optimization $D_{\text{KL}}$, quantization error $L_{\text{quant}}$ and commitment regularization $L_{\text{commit}}$:
\begin{equation}
L_{\text{total}} = L_{\text{recon}} + L_{\text{quant}} + \beta \left( L_{\text{commit}} + D_{\text{KL}}\left(P(z_Q | x) \| P_{\text{prior}} \right) \right)
\end{equation}
where $\beta$ is  a hyperparameter originating from the variational information bottleneck (VIB) objective $J_{IB}=I(z_q;\hat{x}) - \beta I(z_q;x),$ and $D_{KL}$ is the Kullback–Leibler divergence which penalizes deviation from a uniform codeword usage by optimizing the distance between the posterior $P(Z_Q |x)$ and a prior distribution $P_{prior}(Z_Q)$. 
\vspace{-2mm}
\subsection{Transformer: Learning the Autoregressive Prior}
\label{transformers}
After mapping input signals to discrete tokens, a DoT transformer is trained to learn the sequential structure of the latent space statistically. Specifically, it learns the joint probability model of the latent vector of codebook tokens $z_Q$ \cite{ReFormer}.

\subsubsection*{Architecture} The DoT comprises input embedding layer,  positional encodings, and $L$ stacked multi-head attention blocks.
The model processes token sequences beginning with the  Beginning of the Sequence (BOS) token and an  optional class prompt $C$ (see \cite{ReFormer} for more about conditional DoTs), followed by the latent tokens $z_1,z_2,\ldots,z_{d_s-1}$. 
The training loss objective is the negative log-likelihood over the next token prediction:
\begin{equation}
\vspace{-2mm}
L_{\text{CE}} = - \sum_{i=1}^{d_s - 1} \log P(z_{i+1} \mid \text{BOS}, C, z_1, z_2, \ldots, z_i)
\vspace{-1mm}
\end{equation}
Minimizing $L_{CE}$ teaches the model to predict the correct next token at each sequence position, effectively learning the discrete data distribution captured by the VQVAE. In this paper, MONAI \cite{monai2024} DoT is our reference architecture.
\subsubsection*{Auto Regression}
Autoregression is a modeling technique where each output in a sequence is predicted based on previously observed or generated elements. Autoregressive transformers recursively learn the conditional probability distribution of the next token given all prior tokens, which is well-suited for modeling discrete sequences like codebook indices obtained from VQVAE quantization.
During training, the model is exposed to the true previous tokens: 
\begin{equation}
\vspace{-1mm}
P(z_1, z_2, \ldots, z_{d_s}) = \prod_{i=1}^{d_s} P(z_i \mid z^T_1, \ldots, z^T_{i-1}),
\vspace{-2mm}
\end{equation}
where $z^T_i$ is the $i^{th}$ ground-truth token.
At inference time, the model generates new sequences, one token at a time, starting from a special Beginning-of-Sequence (BOS) token. It uses its previously generated outputs to condition the next prediction in a recursive manner:
\begin{equation}
z_1 = \text{BOS}, \quad z_2 \sim P(z_2 \mid z_1), \quad z_3 \sim P(z_3 \mid z_1, z_2), \ldots
\end{equation}
where $z_i$ is the $i^{th}$ token.
\begin{figure}[h]
\centering
\includegraphics[width=0.48\textwidth, height = 4.5cm]{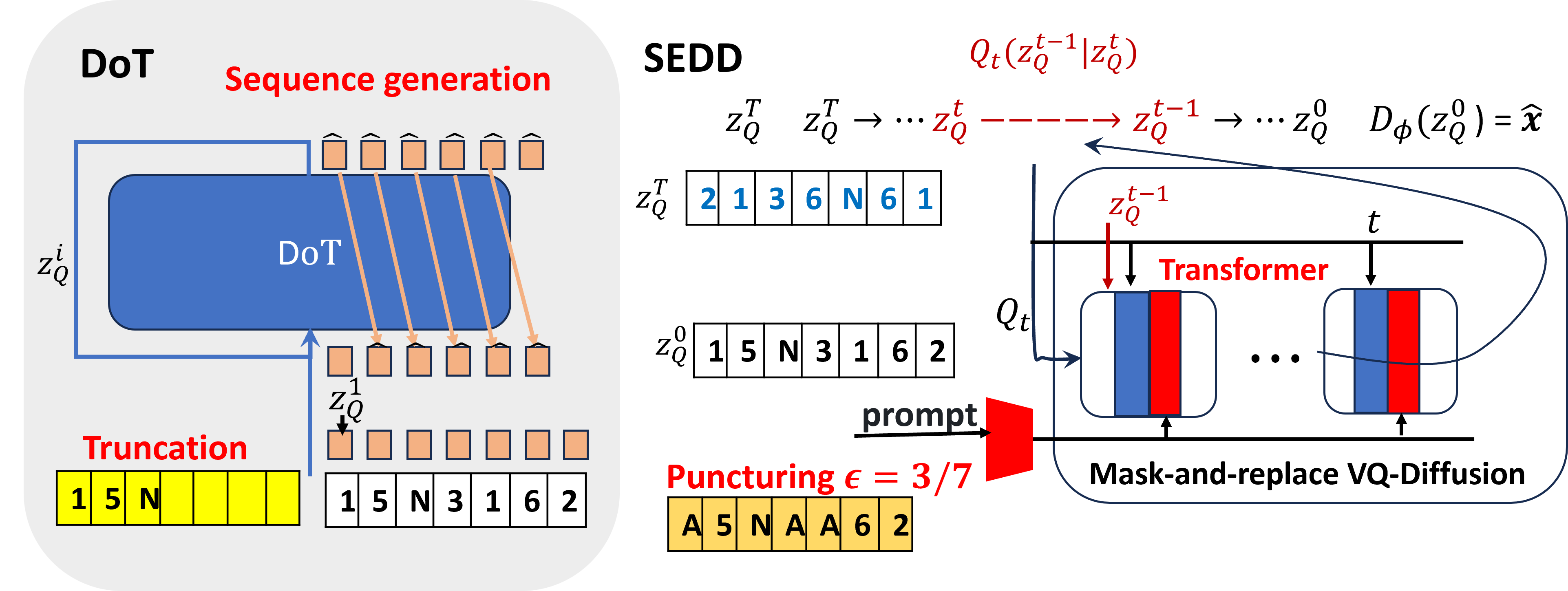}
\vspace{-1mm}
\caption{Two Discernment models: DoT and SEDD differ not only in how they get trained, but also how they generate data, both randomly and from incomplete representations.
}\vspace{-3mm}
\label{fig:vqmodels}
\end{figure}
 This approach enables the DoT to capture complex temporal dependencies and generate coherent sequences of tokens that, when decoded through the VQVAE, reconstruct meaningful signals such as modulated RF waveforms or speech spectrograms. 
\vspace{-2mm}
\subsection{SEDD: Score Entropy Discrete Diffusion}
Discrete diffusion differs from autoregressive models in that it predicts all tokens $z_Q$ simultaneously, enabling faster sampling. However, predicting $z_Q$ involves reverse diffusion process which occurs over a number of  scheduler-iteration steps.  Diffusion schedule guides the training process too, including the processes of adding noise to discrete data and removing it. One way of adding noise is to randomly replace input tokens with a distinguishable mask token $A \notin N_Q$ with varying probability depending on the timestep $t$. 
Next, a suitable neural network model is trained to remove the noise. The SEDD model \cite{lou2024discretediffusionmodelingestimating} is an efficient implementation of the discrete diffusion leveraging  Mask-and-replace (M\&R) strategy. SEDD uses a transformer as the diffusion encoder $s_\theta,$ with full self-attention across all input tokens. 
The model is trained to minimize masked cross-entropy loss:
\begin{equation}
L_{\text{mask-CE}} = - \sum_{i=1}^{d_s} m^t_i \log P(z_i \mid Z_Q, t, C)
\vspace{-2mm}
\end{equation}
where $m^t_i$ is indicating positions of unknown (masked)  tokens at time $t$. The network learns to reconstruct only masked tokens at each step, simplifying the denoising objective.

We next provide more details on how SEDD relates to the mathematical model of diffusion. Discrete diffusion is guided by a Markov chain $p_t\in {\mathbb{R}}^{N}$ s.t. $dp_t/dt=Q_tp_t, p_0\sim p(Z^{tr}_Q).$   $Q_t \in \mathbb{R}^{N\times N}$ are the diffusion matrices for incremental corrupting (over $t\leq T$ steps) of the tokens in the training data $z_Q\in Z^{tr}_Q$.  Transition densities
for each token in $z_Q$ come from the columns of $Q_t;$ Note that $Z^{tr}_Q$  is obtained by mapping the training signal dataset $X^{tr}$ via VQVAE. Hence, $(Q_t)^{d_s}$ is the joint transition probability model which quantifies noise-addition from $z^{t-1}_Q$ to $z^t_Q.$ The reverse process can also be expressed through the matrix $Q_t$ as $dp_{T-t}/dt=\bar{Q}_{T-t}p_{T-t},\ \bar{Q}_t(y,x)=Q_t(x,y)p_t(y)/p_t(x),$ for any token values $y$ and $x\neq y$. The goal of a discrete diffusion model is to construct the
reverse process by learning the ratios $p_t(y)/p_t(x).$ SEDD trains the diffusion encoder to learn this ratio, known as the score: $s_\theta(x, t) \approx \squareb{p_t(y)/p_t(x)}.$

SEDD models the probability of token corruption (forward noise) in the diffusion matrix $Q_t$ differently from the previous discrete diffusion approaches by considering only the transitions resulting in a single-token change per iteration. By doing so,  SEDD reduced the computational complexity of learning by an order of magnitude ((Table~\ref{tab:configs}). 
For the M\&R discrete diffusion strategy, the relationship between the elements of the noise-controlling matrix $Q_t$ (equation~\ref{eq:qt}) is modified from $N\beta_t = 1-\alpha_t$ to 
$N\beta_t = 1-\alpha_t -\gamma_t,$
where $\gamma_t$ quantifies the probability of replacing a token with the mask token $A$, $\alpha_t$ is the probability of keeping the token as is, and $\beta_t$ is the uniform probability of changing the value of the token. Masking makes the corrupted location known and the estimation easier. 

Note that {\em we modified the implementation of SEDD} from \cite{lou2024discretediffusionmodelingestimating} to be able to conditionally model non-language sequences, and most importantly, our noise included masking only, which makes it amenable to the modeling with an erasure channel. This means that $\beta_t=0,$
 resulting in $\alpha_t + \gamma_t=1.$ The forward diffusion is governed by a schedule, either linear or cosine, determining the probability of token masking $\gamma_t$ at each $t$. 
Learning how to denoise is equivalent to learning the generative model of $z_Q$ from pure noise. Additionally, if the technical channel delivers tokens of $z_Q$ out of order (punctured), SEDD can recover it without waiting for all tokens to be delivered.	
\begin{equation}
\begin{bmatrix}
    \alpha_t+\beta_t  & \beta_t & \beta_t  & \dots  & 0 \\
    \beta_t & \alpha_t+\beta_t & \beta_t & \dots  & 0 \\
    \beta_t & \beta_t & \alpha_t+\beta_t & \dots  & 0 \\
    \vdots & \vdots & \vdots & \ddots & \vdots \\
    \gamma_t & \gamma_t & \gamma_t & \dots  & 1
\end{bmatrix}  \label{eq:qt} 
\end{equation}
\subsection{Semantic Channel Pipeline}
The Discernment system employs a two-stage architecture where a VQVAE compresses high-dimensional input signals $x$ into discrete tokens, and the GenAI (a Decoder-only Transformer (DoT) or a SEDD diffusion model) learns the prior distribution over those token sequences. 
At inference (signal synthesis), the GenAI model generates a new token sequence $\widecheck{Z}_Q$ based on the learned prior. 
The generated latent sequence is passed to the VQVAE decoder $D_\theta(\widecheck{z}_Q^c)$ (see Fig.~\ref{fig:vqmodels}), which reconstructs the high-dimensional generated signal (e.g., spectrogram or waveform). 

In the semantic channel framework, the transmitter's signal $x,$ needed for a receiver's task, is processed by VQVAE to obtain $d_s$ discrete tokens which are sent to the receiver. If the tokens $z_Q$ are received over a truncation channel of parameter $\epsilon,$ the receiver obtains the first $(1-\epsilon)d_s$ tokens  (see Fig.~\ref{fig:vqmodels}) and starts the DoT inference process to predict the rest of the tokens. If $z_Q$ is received over a puncturing channel of parameter $\epsilon,$ the receiver acquires $(1-\epsilon)d_s$ tokens and after replacing the erased (missing) tokens with the mask $A,$ starts the SEDD denoising process  ( Fig.~\ref{fig:vqmodels}).

In both cases,  the output of the GenAI at the receiver is passed to the VQVAE decoder $D_\theta(\widecheck{z}_Q)$ to recover the signal in the original domain. This modular and interpretable pipeline separates encoding, latent generative  modeling, and decoding, allowing for easy swapping between generative models and control over conditional (e.g. by erasures) synthesis. 
In this paper, the final synthetic sample is evaluated  by a pre-trained  classifier for various values of $\epsilon.$  
\subsubsection{Classification Models}
To quantitatively assess the quality of generated samples, we train classifiers on the original datasets and evaluate them on synthetic data. 
For Torchsig, using cross-entropy loss, we train a 1D convolutional neural network (CNN) whose input is a 2-channel sequence of length 1024. 
For AudioMNIST, we use a 2D CNN-based classifier tailored to the spectrogram inputs of shape (2, 128, 44). The same cross-entropy loss is used. Classification accuracy on real data samples is 100\% in both cases.  
\vspace{-2mm}
\section{Experimental results}\label{sec:experiments}
\begin{figure} [!htbp]
\vspace{-1mm}
\centering
\includegraphics[width=0.48\textwidth, height = 4.1cm]{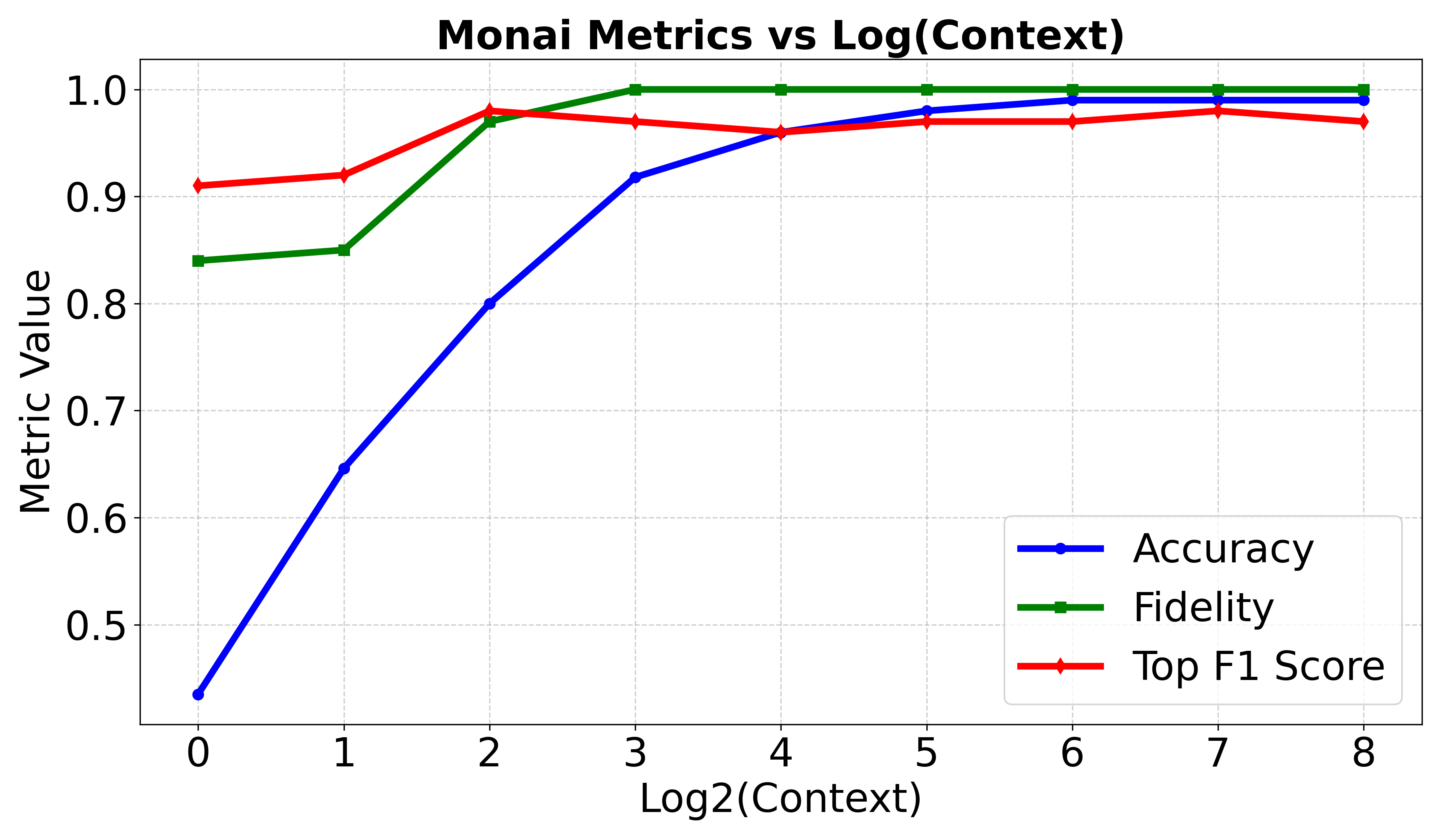}
\caption{Fidelity metrics of the DoT Discernment for the Torchsig classification task as a function of the number of received tokens $t_e=2^x$  out of $d_s=512.$ Shown are the accuracy of classification, statistical data fidelity and top-F1 score of the reconstructed data.} \vspace{-2mm}
\label{fig:truncation}
\end{figure}

Using the baseband signals of 6 common communication waveforms from Torchsig, Fig.~\ref{fig:truncation} presents metrics that illustrate the fidelity of the meaning conveyed by a truncating erasure channel of capacity $C=2^x/d_s$ and the DoT-Discernment. Observe that the $x$ axis goes up to 8, equivalent to $C=0.5.$ i.e., at $x=8,\ \epsilon = 0.5$. Note that Discernment  achieves full accuracy for more severe truncations, at $\epsilon=0.875, C=0.125$ for $x=6$. The top F1 score combines both the fidelity and diversity of the generated samples, and it slightly drops at $x=2,$ as the diversity reduces with the received context length. 

\begin{figure} [!htbp]
\vspace{-3mm}
\centering
\hspace{-2mm}\includegraphics[width=0.49\textwidth, height = 4.6cm]{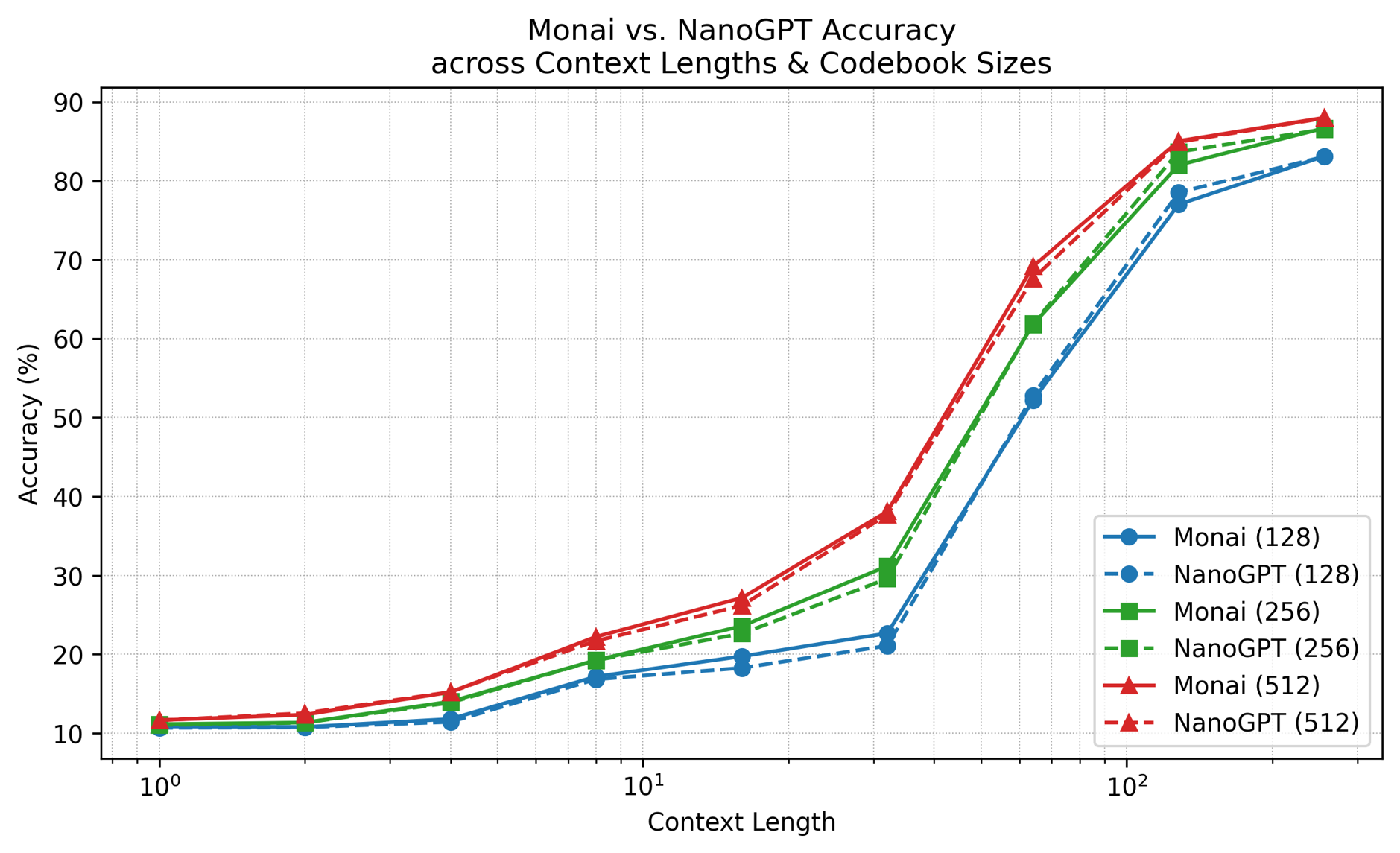}
\caption{Accuracy of the AudioMNIST DoT-Discernement for different received context lengths, different DoT models and different codebook sizes $N$ for the same $\ell,$ $K$ and $d_s$. } \vspace{-2mm}
\label{fig:Audiometrics}
\end{figure}

For the AudioMNIST DoT-Discernement we observed the same behavior across the metrics. In the interest od space, we provide a different perspective on the AudioMNIST in Fig.~\ref{fig:Audiometrics} : we grouped the plots of the context length effect on the accuracy for multiple codebook sizes $N$. All these codebooks, had the same codeword length $\ell$ and $d_s=352.$ Interestingly, the largest $N$ provides the earliest recovery of the task-aligned meaning (accuracy). This is counterintuitive because the largest $N$ provides more uncertainty in terms of the next token as the set $N_Q$ is larger. Future research will address thisissue with more depth and rigor. Full accuracy is achieved for $\epsilon \leq 0.5$. Audio data is less robust to erasures than Torchsig since we compressed it more.

\begin{figure} [!htbp]
\centering
\hspace{-2mm}\includegraphics[width=0.49\textwidth, height = 4.6cm]{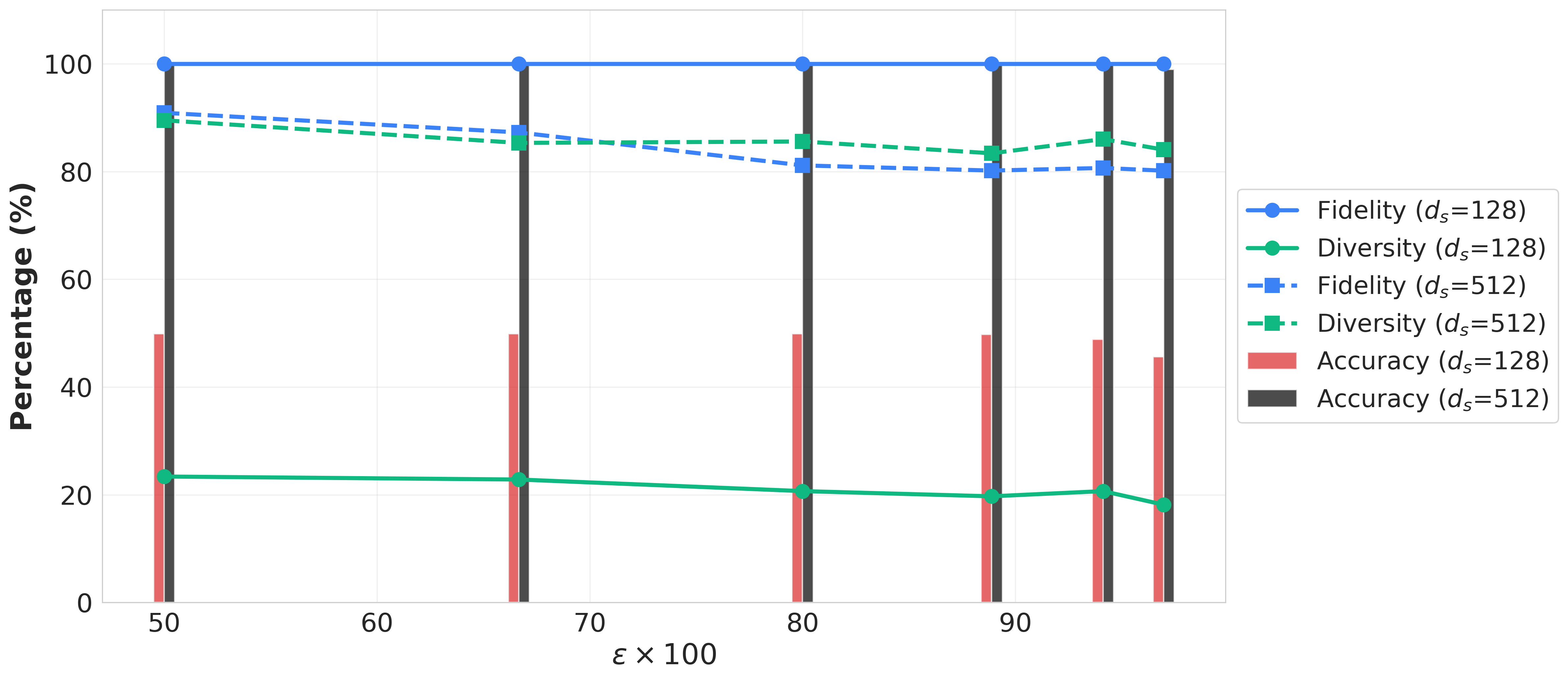}
\caption{Accuracy, fidelity and diversity of the SEDD-Discernement Configuration 1 (N=128) {\bf shows amazing robustness} to increased percentage of lost packets $\epsilon$ ( from 50\% to $\sim$ 97\%) with $d_s=512$ but not with $d_s=128$. Conf. 2 with $\ell=128$ shows the same behavior at $d_s=512$ and $d_s=128$, as does the Conf. 4 for $\ell=256.$ } \vspace{-3mm}
\label{fig:seddmetrics}
\end{figure}

Finally, the results obtained for the SEDD semantic channel show that the task-aligned meaning is only the subject to minor misinterpretation when the majority of tokens are erased, but the overall robustness to token erasure is remarkable, as illustrated in Fig.~\ref{fig:seddmetrics} for $d_s=512 (128).$ This is for the Conf. 1 presented in Table~\ref{tab:configs}. For 11\% of the tokens received, the accuracy drops by only 0.02\% and for 3\% of tokens received ($\epsilon = 0.97$), it drops by 1.07\% (98.93\%). The pattern of token losses is not completely uniform, but it is arbitrary: $\epsilon = 0.5$ is achieved by dropping every other token while $\epsilon=$3\% is achieved by receiving one token after $i=32$ contiguous tokens are dropped. The $\epsilon$ points in between are obtained by dropping i=[2,4,8,16] contiguous tokens after each received token. Note that {\bf in the task of data synthesis}, SEDD achieves the accuracy of 90\% on the data generated  with no context other than the class token, while DoT achieves above 99\%. {\bf In the task of semantic reconstruction}, Fig.~\ref{fig:seddmetrics} shows a more robust semantic system: up to $\epsilon=0.8$, SEDD matches the original data classification accuracy (100\%). Conf. 2 results match Conf.1 for both values of $d_s.$ There are codebook configurations and compression rates that do not achieve this remarkable accuracy, but we omit them due to space limits, and we omit the AudioMNIST SEDD plots as they behave similarly. Details related to DoT and SEDD in their basic generative function are available in \cite{yagnaKMaster}. 
\vspace{-2mm}
\vspace{-5mm}
\vspace{-1mm}
\section{Conclusion}\label{sec:conclude}
Discernment is a semantic communication system that conveys the meaning of physical signals over a technical  communication channel, by leveraging GenAI models in discrete space. Discernment GenAI adapts to the technical channel impairments modeled as erasures by choosing autoregressive or diffusion-based generation algorithms. We demonstrated graceful degradation in both the classification performance and the statistical fidelity of the conveyed meaning for radio and audio signals. The best performance is achieved with discrete diffusion. Future work will examine how Discernment’s robustness to channel impairments scales with latent compression rate, as well as how its computational complexity scales with task complexity. In addition, we plan to incorporate contrastive learning prior to the VQ-VAE to bring semantically similar content closer in latent space, which may further improve completion of partially received latent descriptions.
\vspace{-4mm}
\bibliographystyle{IEEEtran}
\bibliography{references}
\end{document}